\documentclass[]{my_aastex701}

\usepackage{libertine}
\usepackage{libertinust1math}
\usepackage[T1]{fontenc}

\shorttitle{The Calm before the Storm}
\shortauthors{Denker et al.}
\received{24 November 2025}
\submitjournal{Research Notes of the American Astronomical Society}

\begin{document}


\title{\large The Calm before the Storm:\\
    High Spatial Resolution Mosaic of Active Region NOAA 14274 at the Onset of an X1.2 Flare}

\author[orcid=0000-0002-7729-6415, sname='Denker']{Carsten Denker}
\affiliation{Leibniz Institute for Astrophysics Potsdam (AIP), 
    An der Sternwarte 16, 14482 Potsdam, Germany}
\email{cdenker@aip.de}  

\author[orcid=0000-0003-2059-585X, sname='Kamlah']{Robert Kamlah}
\affiliation{Leibniz Institute for Astrophysics Potsdam (AIP), 
    An der Sternwarte 16, 14482 Potsdam, Germany}
\email{rkamlah@aip.de}  

\author[orcid=0000-0003-1054-766X, sname='Verma']{Meetu Verma}
\affiliation{Leibniz Institute for Astrophysics Potsdam (AIP), 
    An der Sternwarte 16, 14482 Potsdam, Germany}
\email{\textcolor{xlinkcolor}{mverma@aip.de}}  

\author[orcid=0000-0002-0484-7634, sname='Pietrow']{Alexander G.\ M.\ Pietrow}
\affiliation{Leibniz Institute for Astrophysics Potsdam (AIP), 
    An der Sternwarte 16, 14482 Potsdam, Germany}
\email{apietrow@aip.de}


\begin{abstract}
Active region NOAA~14274 produced some of the strongest flares of Solar Cycle~25, including the X1.2 and X5.1 flares on 10 and 11 November 2025, respectively. We present the first large mosaic of speckle-restored images obtained with the improved High-resolution Fast Imager (HiFI+) at the 1.5-meter GREGOR solar telescope at the Observatorio del Teide in Iza\~na, Tenerife, Spain. The observations were obtained approximately 30~minutes before the onset of the X1.2 flare. The active region exhibited strongly curved penumbral filaments, sunspot rotation, and shear motions along the polarity inversion line (PIL), which led to a highly stressed magnetic field configuration that stored sufficient energy to release multiple M- and X-class flares. The first flare signatures appeared as small-scale brightenings, each with a width of a few tenths of an arcsecond, that trace penumbral filaments in the trailing sunspot.
\end{abstract}


\renewcommand{\baselinestretch}{1.15}\selectfont


\section{Introduction} 

Solar telescopes with large apertures have a relatively small field of view (FOV), typically around 100\arcsec\ or less. This is generally too small to cover entire active regions \citep[see][for an exception]{Kamlah2025}. Raster scans can create a larger FOV, but this comes at the cost of temporal resolution \citep{DeWilde2025}. This is especially true if many images must be recorded for image restoration \citep[e.g.,][]{Woeger2021} or if integral field units \citep[e.g., ][]{DominguezTagle2022} are used to record spectropolarimetric data. \citet{Kuckein2016} presented a mosaic of slit-reconstructed spectroheliograms obtained with the echelle spectrograph of the 0.7-meter Vacuum Tower Telescope \citep[VTT,][]{vonderLuehe1998}, which covered an 800\arcsec-long H$\alpha$ filament. Mosaics have also been used to study the center-to-limb variation (CLV) of spectral lines \citep{Pietrow2023}. The CLV can also be assessed with a full-disk instrument, such as Challan \citep{Yang2025}, or with a high spatial and spectral resolution spectrograph \citep{Verma2025}, using drift scans. Drift scans can be considered an extreme form of raster scans.


\section{Observations} 

\begin{figure}
\centering
\includegraphics[width=\textwidth]{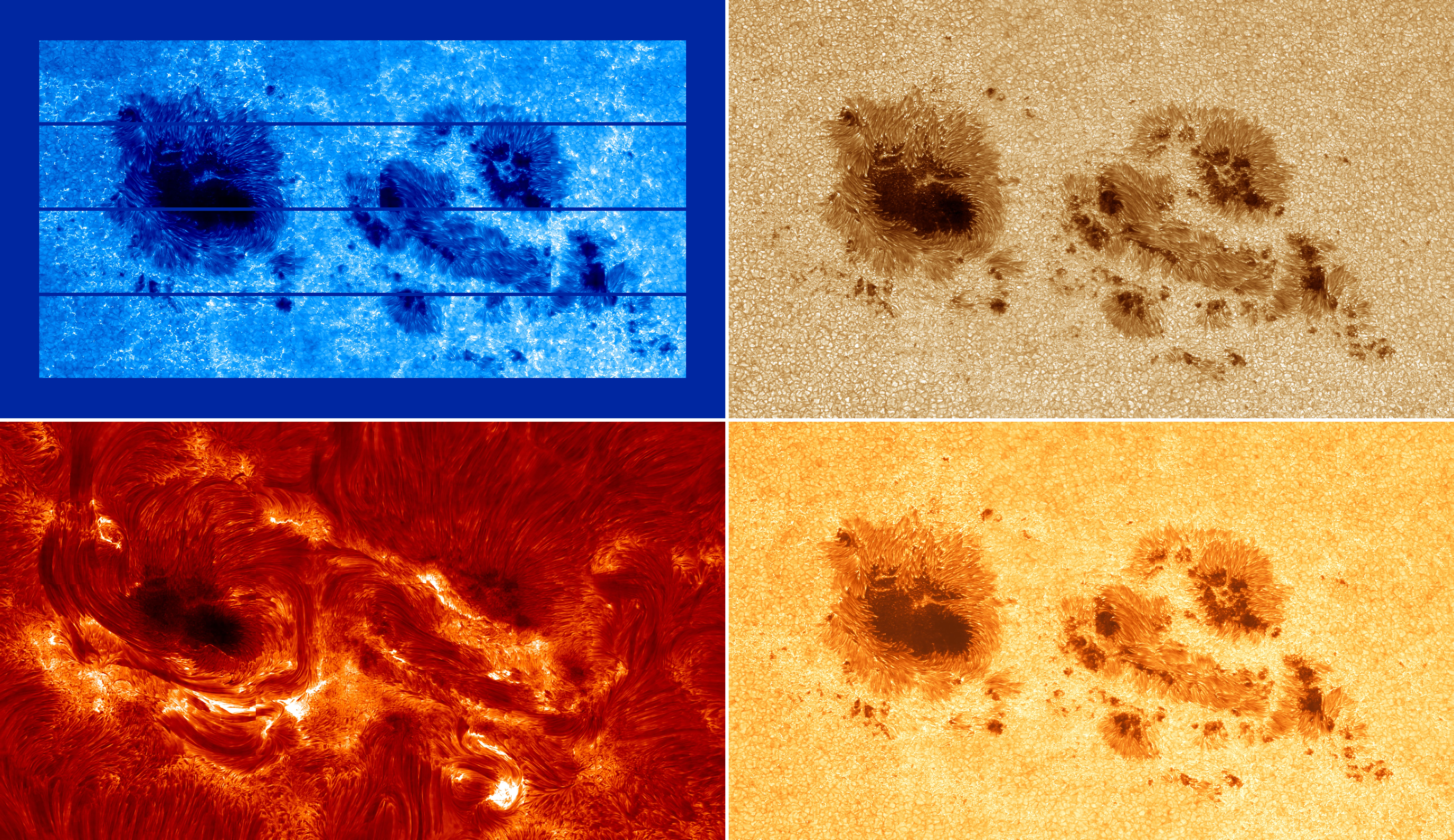}
\caption{The mosaics of speckle-restored images have a combined FOV of $240\arcsec \times 
    150\arcsec$ and shows active region NOAA~14274 at around 08:33~UT on 10~November 2025, about 30~min before an X1.2 flare began. The mosaics were created using four HiFI+ channels: Ca\,\textsc{ii}\,H $\lambda$396.8~nm, TiO $\lambda$705.8~nm, and broad- and narrow-band H$\alpha$ $\lambda$656.3~nm (\textit{clockwise from the top-left panel}). The FOV of the Ca\,\textsc{ii}\,H channel ($48.2\arcsec \times 30.8\arcsec$) is too small to overlap vertically (as shown by the horizontal blue lines in the Ca\,\textsc{ii}\,H panel), while the FOV of the other channels ($76.5\arcsec \times 60.5\arcsec$) is much larger.}
\label{RASTER}
\end{figure}

The observations were carried out on 10~November 2025 with the 1.5-meter GREGOR telescope \citep{Schmidt2012}. Images with a high cadence of 100~Hz and a high spatial resolution of 0.025\arcsec/0.05\arcsec\ pixel$^{-1}$ were recorded in two channels of HiFI+ \citep{Denker2018, Denker2023}. A raster scan with a 30\arcsec-step size produced 28 telescope pointings, and three datasets were acquired with 500 images each, resulting in $7 \times 4 \times 3 = 84$ datasets. The 100 best images per dataset were selected for image restoration using the Kiepenheuer Institute Speckle Imaging Package \citep[KISIP,][]{Woeger2008}, which is accessible via the sTools data processing pipeline developed for HiFI+ \citep{Kuckein2017}. The best restored image for each pointing was selected to create a mosaic (Figure~\ref{RASTER}), resulting in a 30-second cadence for each raster step and a total time of 14~min to record the entire mosaic. 

\section{Results and Conclusions} 

The active region NOAA~14274, which produced numerous flares, was observed 4\,--\,11 November 2025, using the HiFI+ instrument and the GREGOR infrared spectrograph \citep[GRIS,][]{Collados2012}. During the two days with the two X-class flares, nearly 10\,000 speckle restoration datasets were recorded with HiFI+, and multiple spatial scans were obtained with GRIS. On November 10, the $\beta\gamma\delta$-region was still growing and had reached a sunspot area of 900 millionths of the Sun’s visible hemisphere when it produced an X1.2-class flare. The flare started at 08:55:00~UT and reached its peak at 09:19:00~UT. The penumbral filaments of the trailing spot are strongly curved and nearly tangential in the southern part of the sunspot \citep{Lim2013}, where multiple parasitic magnetic polarities are present within or near the penumbra. The leading part of the active region is characterized by braided penumbral filaments that give the impression of submerging filament bundles. Shear motions along the PIL and sunspot rotation continuously stress the magnetic field topology \citep{Wang2015, Verma2018}, providing sufficient energy to power multiple M- and X-class flares. First, flare signatures appear in the western part of the penumbra of the trailing sunspot. Next, they appear along the filament wrapping around the trailing sunspot in the south, where H$\alpha$ line-core brightenings are already visible 30~min before the flare. Then, the northern leading part of the region flares up. Most of the activity occurs in the central part of the region where the flare ribbons extend into the umbra of the trailing sunspot and the H$\alpha$ moss region to the south. The first post-flare loops formed around 09:17~UT, connecting the moss region and the trailing sunspot. The data presented in this research note provide an initial look at the data quality during the November 2025 GREGOR observing campaign.


\section*{Data Availability}

\noindent The HiFI+ data are in general embargoed for one year after which they become publicly available (\href{https:\\gregor.aip.de}{gregor.aip.de}). 

\begin{acknowledgments}
The 1.5-meter GREGOR solar telescope was built by a German consortium under the leadership of the Institute for Solar Physics (KIS) in Freiburg with the Leibniz Institute for Astrophysics Potsdam (AIP), the Institute for Astrophysics Göttingen (IAG), and the Max Planck Institute for Solar System Research (MPS) in Göttingen as partners, and with contributions by the Instituto de Astrofísica de Canarias (IAC) and the Astronomical Institute of the Academy of Sciences of the Czech Republic (ASU). This research has made use of the bibliographic services of NASA's Astrophysics Data System (ADS). AP was supported by grant PI~2102/1-1 from the Deutsche Forschungsgemeinschaft (DFG). MV acknowledges the support by a WISER grant from the Indo-German Science \& Technology Center (IGSTC-05373). 
\end{acknowledgments}

\setlength{\parindent}{0em}
\facilities{GREGOR solar telescope \citep{Schmidt2012} -- HiFI+ \citep{Denker2018, Denker2023}}
\software{sTools \citep{Kuckein2017} -- KISIP \citep{Woeger2008}}

\renewcommand{\baselinestretch}{1.0}\selectfont


\newcommand{\BIBdecl}{\setlength{\itemsep}{1em}}
\bibliographystyle{spr-mp-sola}
\bibliography{apj-jour, hifiplus_raster}

\end{document}